\documentclass[10pt,a4]{article}  %>>> use for US letter paper
\def\ra{\rangle}
\def\la{\langle}
\def\bege{\begin{equation}}
\def\ende{\end{equation}}
\def\begarr{\begin{eqnarray}}
\def\endarr{\end{eqnarray}}
\def\no{\noindent}
\def\non{\nonumber}

\def\unity{\mbox{\small 1} \!\! \mbox{1}}

\usepackage{graphicx}

\title{Quantum Interferometric Sensors} 

\author{Kishore T. Kapale$^1$, Leo D. Didomenico$^2$, Hwang Lee$^3$, \\
Pieter Kok$^4$, and Jonathan P. Dowling$^{3,5}$\\
 $^1$\small \it Quantum Computing Technologies Group, Jet Propulsion Laboratory,\\
\small \it California Institute of Technology,
Mail Stop 126-347,\\
\small \it 4800 Oak Grove Drive, Pasadena, California 91109-8099\\
$^2$\small \it  Extreme Energetics, 
Lawrence Livermore National Laboratory, \\
\small \it 7000 East Avenue
Building 3180 Suite 117
Livermore CA, 94550
\\
$^3$\small \it Hearne Institute for Theoretical Physics, Department of Physics \& Astronomy,\\
\small \it Louisiana State University, Baton Rouge, Louisiana 70803-4001\\
$^4$\small \it Quantum Information Processing Group, Hewlett-Packard Laboratories,\\ \small \it Filton Road, Stoke Gifford,  Bristol BS348QZ, United Kingdom\\
$^5$\small \it Institute for Quantum Studies, Department of Physics, \\ \small \it Texas A\&M  University, College Station, Texas 77843}
\begin{document}
\begin{titlepage}
%\begin{center}
\vskip-5cm
\maketitle 
%\date{\today}
\begin{abstract}
Quantum entanglement has the potential to revolutionize 
the entire field of interferometric sensing by providing many orders 
of magnitude improvement in interferometer sensitivity.
The quantum-entangled particle interferometer approach is 
very general and applies to many types of interferometers. 
In particular, without nonlocal entanglement, a generic classical 
interferometer has a statistical-sampling shot-noise limited sensitivity 
that scales like $1/\sqrt{N}$, where $N$ is the number of particles passing 
through the interferometer per unit time. 
However, if carefully prepared quantum correlations are engineered between the particles, 
then the interferometer sensitivity improves by a factor of $\sqrt{N}$  
to scale like $1/N$, which is the limit imposed by the Heisenberg Uncertainty Principle. 
For optical interferometers operating at milliwatts of optical power, 
this quantum sensitivity boost corresponds to 
an eight-order-of-magnitude improvement of signal to noise. 
This effect can translate into a tremendous science pay-off for space missions. 
For example, one application of this new effect is to fiber optical gyroscopes 
for deep-space inertial guidance and tests of General Relativity (Gravity Probe B). 
Another application is to ground and orbiting optical interferometers for gravity wave detection, Laser Interferometer Gravity Observatory (LIGO) 
and the European Laser Interferometer Space Antenna (LISA), respectively. 
Other applications are to Satellite-to-Satellite laser Interferometry (SSI) 
proposed for the next generation Gravity Recovery And Climate Experiment (GRACE II). 
\end{abstract}
%\end{center}

\end{titlepage}

%%%%%%%%%%%%%%%%%%%%%%%%%%%%%%%%%%%%%%%%%%%%%%%%%%%%%%%%%%%%%

\section{Introduction}
\label{sect:intro}  % \label{} allows reference to this section

In a conventional optical interferometer, 
in which light in a coherent state enters via only one port, 
that the phase sensitivity scales as $1/\sqrt{\bar n}$, 
where ${\bar n}$ is the mean number of photons to 
have passed through the interferometer in an integration time \cite{scully97}.
It would seem that any desired sensitivity could be attained by 
simply increasing the laser power and hence ${\bar n}$. 
However, since the phase sensitivity scales only slowly as $1/\sqrt{\bar n}$, 
the laser power rapidly becomes so large that 
the power fluctuations at the interferometer's mirrors introduce 
additional noise terms that eventually limit the device's overall sensitivity. 
Steady improvements in optical laser gyroscope designs indicate 
that quantum noise fluctuations such as these will be the dominant effect 
limiting laser gyroscope accuracy in the near future. 
Much of the early interest in coherent photon-state squeezing centered 
on overcoming this signal-to-noise roadblock. 
In fact, much of the early work on photon squeezing was motivated 
by the goal of improving this power law in the sensitivity scaling 
in order to make space-borne, optical interferometric gravity wave detectors.

In 1981 Caves showed that when phase-squeezed coherent states are 
fed into both input ports of the interferometer, 
then phase sensitivity can asymptotically approach  $1/{\bar n}$
for large mean photon number  ${\bar n}$, 
which is proportional to the optical input power \cite{caves81}. 
This is a great achievement in that the total laser power required for 
a given amount of phase sensitivity is greatly reduced. 
For a typical milliwatt laser gyro or optical interferometer gravity wave detector, 
this amounts to about an eight order-of-magnitude increase in rotation 
or phase-shift sensitivity of the interferometer from the quadratic increase 
in the power law alone. 
On the other hand, in 1986, Yurke
as well as Yuen had considered the question of 
phase noise reduction using correlated quantum particles in number states -- rather than squeezed coherent states -- also incident upon both input ports of 
a Mach-Zehnder interferometer in a highly quantum-entangled fashion \cite{yurke86a,yuen86}. 
For quantum number states -- unlike squeezed coherent states -- there are no number fluctuations. 
This rules out squeezing in the conventional sense. 
Nevertheless, Yurke was able to show that if $N$ particles entered 
into each input port of the interferometer in nearly equal numbers 
-- and in a highly correlated and entangled fashion -- then it was 
indeed possible to obtain the desired asymptotic Heisenberg-limited 
phase sensitivity scaling of order  $1/N$ for large $N$. 
This should be compared to the scaling of  $1/\sqrt{N}$ that is the best 
one can do using uncorrelated particles or only one input port.

Shortly after Yurke's first paper was published, 
there appeared a second, related paper by Yurke, McCall, and Klauder, 
indicating how such a Heisenberg-limited ($1/N$) interferometer 
might be obtained using correlated photons emanating from a nonlinear, 
optical four-wave mixing device \cite{yurke86b}. 
Since then, there have been a large number of seemingly independent papers suggesting 
related ideas of using correlated number-state photons to make an improved interferometer, 
using a wide variety of nonlinear optical devices for correlated photon generation, 
and different choices of entangled photon input states. 
A recent paper by Hall and co-workers suggested that 
the requisite entangled photons can be produced at high power 
in a nonlinear optical parametric oscillator, 
and then sent into both input ports of a Mach-Zehnder interferometer 
to achieve the desired increase interferometric gyroscopic sensitivity, 
approaching the Heisenberg limited quadratically-increased power law of $1/N$ \cite{kim98}.
  
A similar improvement in measurement sensitivity can be achieved in
the determination of frequency standards and spectroscopy. 
Wine\-land and co-workers first showed that the best possible precision in frequency
standard is obtained by using maximally entangled states
\cite{wineland96}. Similarly, it was shown that
this improved sensitivity can be exploited in atom-laser gyroscopes
\cite{dowling98}. 
In later sections we describe general features in interferometric sensors
and their quantum enhancement.

\section{Quantum limit in phase estimation}

\begin{figure}
   \begin{center}
   \begin{tabular}{c}
   \includegraphics[height=3cm]{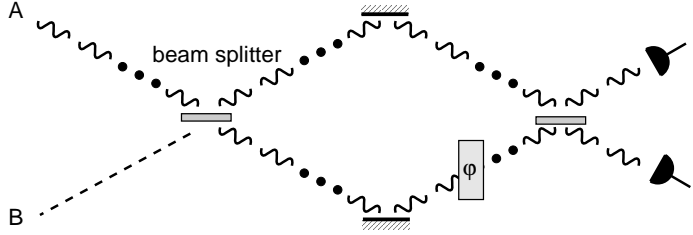}
   \end{tabular}
   \end{center}
   \caption
{\label{mz-1}
An typical Mach-Zehnder interferometer
where the input light is incident on input port A while only
vacuum comes into input port B.
}
\end{figure}

Consider an ensemble of $N$ two-state systems in the state:
\begarr
|\varphi\rangle = {1 \over \sqrt{2}}
(|0\rangle+e^{i\varphi}|1\rangle)
\endarr
\no
where $|0\rangle$ and $|1\rangle$ denote the two basis states.
In a Mach-Zehnder interferometer, the input light field is divided
into two different paths by a beam splitter, and recombined by another
beam splitter. The phase difference between the two paths is then
measured by balanced detection of the two output modes (see
Fig.~\ref{mz-1}).
We may think of the upper and lower paths as the two states
in which a single light quanta can occupy.   
Then, if the photons enter only into the port A, 
the input state can be represented as $|0\ra$.
After the 50/50 beam splitter this transforms to
$(|0\rangle+|1\rangle)/\sqrt{2}$ (up to a certain phase).
And the quantum state $|\varphi\rangle$ now
represents the single photon state after the phase shifter.

The phase information can be obtained by measurement of an observable
$\hat{A} =|0\rangle\langle 1| + |1\rangle\langle 0|$. The expectation
value of $\hat{A}$ is then given by
\begin{equation}
  \langle\varphi|\hat{A}|\varphi\rangle=\cos\varphi\; .
\end{equation}
\no
When we repeat this experiment $N$ times, we obtain
%\be
 $ \langle\varphi_R|\hat{A}_R|\varphi_R\rangle=N\cos\varphi,
$
%\ee
where
$
|\varphi_R\rangle = |\varphi\rangle_1 \ldots |\varphi\rangle_N
$,
and 
$\hat{A}_R = {\mbox{\Large $\oplus$}}_{k=1}^N 
  \hat{A}^{(k)}
$.
Since $\hat{A}_R^2=\unity$, the variance of $\hat{A}_R$, given $N$
samples, is readily computed to be $(\Delta A_R)^2 =
N(1-\cos^2\varphi) = N \sin^2 \varphi$. According to estimation
theory\cite{helstrom76}, we have  
\begin{equation}\label{est}
  \Delta\varphi_{\rm SL} = \frac{\Delta A_R}{|d\langle
  \hat{A}_R\rangle/d\varphi|} = \frac{1}{\sqrt{N}}\; .
\end{equation}
This is the standard variance in the parameter $\varphi$ after $N$
trials. In other words, the uncertainty in the phase is inversely
proportional to the square root of the number of trials. 
This is called the {\em shot-noise limit}.

With a coherent laser field as the input the phase
sensitivity is given by the shot noise limit $N^{-\frac{1}{2}}$, where
$N$ is the average number of photons passing though the interferometer
during measurement time. When the number of photons is exactly known
(i.e., the input is a Fock state $|N\rangle$), the phase sensitivity
is still given by $N^{-\frac{1}{2}}$, indicating that the photon
counting noise does not originate from the intensity fluctuations of
the input beam, but rather from the Poissonian ``sorting noise'' of
the beam splitter \cite{dowling98}.

Now consider a maximally entangled ``N00N'' state
\begin{equation}\label{entang}
 |\varphi_N\rangle\equiv {1 \over \sqrt{2}}
 |N,0\rangle + e^{iN\varphi}|0,N\rangle\; , 
\end{equation}
where $|N,0\rangle$ and $|0,N\rangle$ are 
collective states of $N$ particles, defined as
\begarr
 |N,0\ra &=& |0\ra_1 |0\ra_2 \cdot\cdot\cdot |0\ra_N \non \\
 |0,N\ra &=& |1\ra_1 |1\ra_2 \cdot\cdot\cdot |1\ra_N .
 \label{noon}
\endarr
The relative phase $e^{iN\varphi}$ is accumulated when each particle
in state $|1\ra$ acquires  a phase shift of $e^{i \varphi}$. 
An important question now is---what do we need to measure in order to extract
the phase information? 
Recalling the single-particle case of ${\hat
  A}=|0\ra\la 1| + |1\ra \la 0|$, we need an observable that does what
the operator $|0,N\rangle\langle N,0| + |N,0\rangle\langle 0,N|$
does. For the given state of Eq.~(\ref{entang}), we can see that this
can be achieved by an observable, 
$\hat{A}_N = {\mbox{\Large $\otimes$}}_{k=1}^N \hat{A}^{(k)} $. 
The expectation value of
$\hat{A}_N$ is then 
\begin{equation}\label{cosn}
  \langle\varphi_N |\hat{A}_N| \varphi_N\rangle = \cos N\varphi\; ,
\end{equation}
where the $N$-fold increase in oscillation frequency is the origin of the quantum lithography effect---discussed in Sec.\ 5, below.
Again, $\hat{A}_N^2=\unity$, and $(\Delta A_N)^2 = 1-\cos^2 N\varphi =
\sin^2 N\varphi$. Using Eq.\ (\ref{est}) again, we obtain the
so-called {\em Heisenberg limit} (HL) of the minimal detectable phase:  
\begin{equation}\label{bol}
  \Delta\varphi_{\rm HL} = \frac{\Delta A_N}{|d\langle \hat{A}_N
  \rangle/d\varphi|}=\frac{1}{N}\; .
\end{equation}
The precision in $\varphi$ is increased by a factor $\sqrt{N}$ over
the standard shot-noise limit. Of course, the preparation of a quantum
state such as Eq.~(\ref{entang}) is essential to the given
protocol \cite{wineland96}.

\section{Quantum Rosetta stone}

\begin{figure}
   \begin{center}
   \begin{tabular}{c}
   \includegraphics[height=9cm]{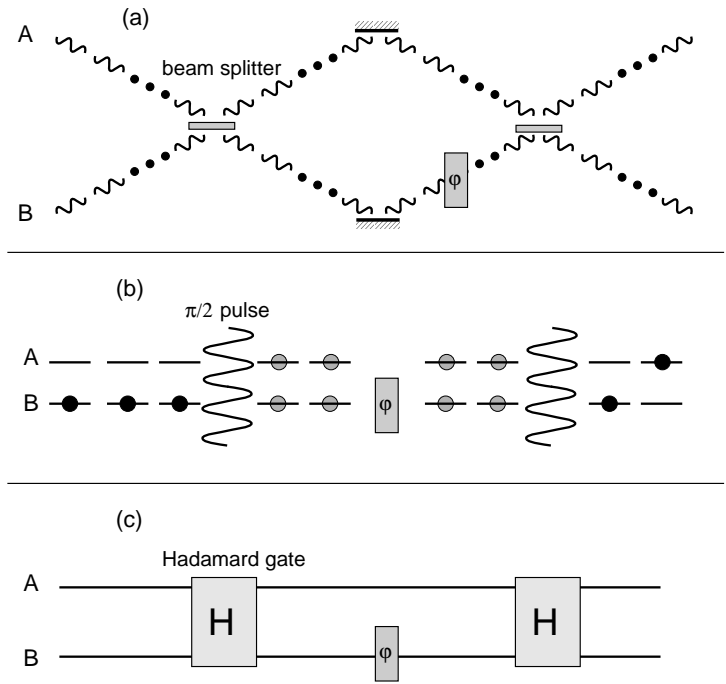}
   \end{tabular}
   \end{center}
   \caption
{\label{fig-1}
Three distinct
representations of a phase measurement:
(a) a Mach-Zehnder interferometer,
(b) Ramsey spectroscope,
and (c) a generic quantum logic gate.
The two basis states of a qubit,
$|0\ra$ and $|1\ra$, may be regarded as the atomic
two levels, or the two paths in a Mach-Zehnder interferometer.
The state $|\varphi\rangle$ can be regarded as
a single photon state just before the second beam splitter 
in the Mach-Zehnder interferometer, or the single atom state
just before the last $\pi/2$-pulse in the Ramsey interferometer.
}
\end{figure}

In a Ramsey spectroscope, atoms are put in a superposition of the
ground state and an excited state with a $\pi/2$-pulse
(Fig.~\ref{fig-1}b). After a time interval of free evolution, a second
$\pi/2$-pulse is applied to the atom and, depending on the relative
phase shift obtained by the excited state in the free evolution, the
outgoing atom is measured either in the ground or the excited state. 
Repeating this procedure $N$ times determines the phase $\varphi$ 
with precision $1/\sqrt{N}$. 
This is essentially an atomic 
clock. Both procedures, the optical Mach-Zehnder interferometer and
the Ramsey spectroscope, are methods to measure the phase shift, either
due to the path difference in the interferometer, or to the
free-evolution time in the spectroscope. When we use entangled atoms
in the spectroscope, we can again increase the sensitivity of the
apparatus.

A similar situation can be found  in a quantum circuit where
a qubit that undergoes a Hadamard transform
$H$, then picks up a relative phase and is then transformed back with
a second Hadamard transformation (Fig.~\ref{fig-1}c). This
representation is more mathematical than the previous two, and it
allows us to extract the unifying mathematical principle that connects
the three systems. In all protocols, the initial state is transformed
by a discrete Fourier transform (beam splitter, $\pi/2$-pulse or
Hadamard), then picks up a relative phase, and is transformed back
again. 
This is the standard quantum finite Fourier transform, such as used 
in the implementation of Shor's algorithm \cite{ekert96}. 
It is {\em not} the same as the 
classical fast-finite algorithm in engineering---it is exponentially faster.

As a consequence the phase shift (which is hard to measure
directly) is applied to the transformed basis. The result is a bit flip
in the initial, {\em computational}, basis $\{ |0\rangle,|1\rangle\}$,
and this is readily measured. We call the formal analogy between these
three systems the {\em quantum Rosetta stone} \cite{lee02b}.
These schemes can be generalized from measuring a simple phase shift to
evaluating the action of a unitary transformation $U_f$ associated
with a complicated function $f$ on multiple qubits. 
Such an evaluation is also
known as a quantum computation. 
The concept 
of quantum computers is therefore to exploit quantum interference in
obtaining the outcome of a computation of $f$. In this light, a
quantum computer is nothing but a complicated multiparticle quantum
interferometer \cite{ekert98}.

This logic may also be reversed:
a quantum interferometer is therefore a simple quantum computer.
Often, when dis\-cuss\-ing, 
Heisen\-berg-limited interferometry
with complicated entangled states,
one encounters the critique that such states are
highly susceptible to noise and even one or
a few uncontrolled interactions with the environment
will cause sufficient degradation of the device and
recover only the shot-noise limit.
Apply the {\em quantum Rosetta stone} by replacing the term
``quantum interferometer'' with ``quantum computer''
and we recognize the exact same critique that
has been leveled against quantum computers for years.
However, for quantum computers, we know the response---to apply
quantum error-correcting techniques and encode in decoherence
free subspaces.
Quantum interferometry is just as hard (or easy) as quantum computing!
The same error-correcting tools that we believe
will make quantum computing a reality,
will also be enabling for quantum interferometry.

\section{Heisenberg-limited interferomery}

There have been various proposals for achieving Heisenberg-limited
sensitivity, corresponding to different physical realizations of the
state $|\varphi_N\rangle$ and observable $\hat{A}_N$ in Eq.~(\ref{cosn}).
Here, we discuss three different approaches, categorized according to
the different quantum states.

\subsection{Yurke states}\label{yurke-sec}

By utilizing the $su(2)$ algebra of spin angular momentum, 
in 1986 it was shown that  \cite{yurke86a,yuen86,yurke86b}
with a suitably correlated input state
the phase sensitivity can be improved to $1/N$. 
Let ${\hat a}^\dagger$, ${\hat b}^\dagger$ denote the 
creation operators for the two input modes in Fig.~\ref{fig-1}a. 
In the Schwinger representation, 
the common eigenstates of ${\hat J}^2$ and ${\hat J}_z$ 
are the two-mode Fock states $|j,m\ra = |j+m\ra_A |j-m\ra_B$,
where 
\begarr
 {\hat J}_x & = &( {\hat a}^\dagger {\hat b} + {\hat b} {\hat
   a}^\dagger )/2\; ; \quad {\hat J}_y = -i( {\hat a}^\dagger {\hat b} -
 {\hat b} {\hat a}^\dagger )/2\; ; \non \\
 {\hat J}_z & = &( {\hat a}^\dagger {\hat a} - {\hat b}^\dagger {\hat
   b} )/2\; ; \quad {\hat J}^2 = {\hat J}_x^2 + {\hat J}_y^2 + {\hat
  J}_z^2\; .
\label{jm}
\endarr
The interferometer can be described by the rotation of the angular
momentum vector, where ${\hat a}^\dagger {\hat a} + {\hat b}^\dagger 
{\hat b}= N = 2j$, and the 50/50 beam splitters and the phase shift
are corresponding to the operators $e^{i \pi {\hat J}_x/2}$ and
$e^{i\varphi {\hat J}_z}$, respectively.  

For spin-1/2 fermions, the entangled input state (which we call the
`Yurke state') $|\varphi_N\ra_{\rm Y}$ is given by
\begarr
 |\varphi_N\ra_{\rm Y} &=& {1 \over \sqrt{2}} \left[ \left|j=\mbox{$\frac{N}{2}$}, m
     =\mbox{$\frac{1}{2}$}\right\ra + \left|j=\mbox{$\frac{N}{2}$},
     m=-\mbox{$\frac{1}{2}$} \right\ra \right] \non \\ 
 &=& {1 \over \sqrt{2}} \left[
   \left|\mbox{$\frac{N+1}{2}$},\mbox{$\frac{N-1}{2}$} 
   \right\ra_{AB} +
   \left|\mbox{$\frac{N-1}{2}$},\mbox{$\frac{N+1}{2}$} \right\ra_{AB} 
 %\left|{N +1\over 2}\right\ra_A \left|{N -1 \over 2} \right\ra_B
 %+ \left|{N -1 \over 2}\right\ra_A \left| {N +1 \over 2} \right\ra_B
 \right],
\label{yurke}
\endarr

\no
where the notion of $|j,m\ra$ follows the definition given in
Eq.~(\ref{jm}) and the subscripts $AB$ denote the two input modes. For
bosons, a similar input state, namely $|j=N/2,m=0\ra + |j=N/2,m=1\ra$,
has been proposed \cite{yurke86b,yuen86}. The measured observable $\hat{A}_N$
is given by ${\hat J}_z$. After evolving the state $|\varphi_N\ra_{\rm Y}$
(in the Ramsey spectroscope or the Mach-Zehnder interferometer with
phase shift $\varphi$), the phase sensitivity $\Delta\varphi$ can be
determined to be proportional to $1/N$ for special values of
$\varphi$.
Although the input state of Eq.~(\ref{yurke}) was proposed
for spin-1/2 fermions, the same state with bosons also yields the
order of $1/N$
phase sensitivity \cite{dowling98}.

\subsection{Dual Fock states}
In 1993 Holland and
Burnett proposed the use of so-called {\em dual Fock states}
$|N\rangle_A \otimes |N\rangle_B$ for two input modes $A$
and $B$ of the Mach-Zehnder interferometer 
in order to achieve Heisenberg-limited sensitivity \cite{holland93}. 
Such a
state can be generated, for example, by degenerate parametric down
conversion, or by optical parametric oscillation \cite{kim98}. 

To obtain increased sensitivity with dual Fock states, 
some special detection scheme is needed.
In a conventional Mach-Zehnder
interferometer only the difference of the number of photons at the
output is measured. 
Similarly, in atom interferometers, measurements are performed by
counting the number of atoms in a specific internal state using
fluorescence.
For the schemes using dual Fock-state input, an
additional measurement is required since the average in the intensity
difference of the two output ports does not contain information about
the phase shift. 
One measures both the sum and the difference
of the photon number in the two output modes \cite{holland93}. 
The sum contains information about the total photon number, 
and the difference contains
information about the phase shift. Information about the total photon
number then allows for post-processing the information about the
photon-number difference.
A combination of a direct measurement of the variance
of the difference current and a data-processing method based on
Bayesian analysis was proposed \cite{kim98}.

For atom interferometers a quantum nondemolition measurement is
required to give the total number of atoms \cite{bouyer97}. In a
similar context, Yamamoto and co-workers devised an atom interferometry
scheme that uses a squeezed $\pi/2$ pulse for the readout of the input state
correlation \cite{yamamoto95}. 

Due to its simple form, the dual Fock-state approach sheds new
light on Heisenberg-limited interferometry. In particular, exploiting
the fact that atoms in a Bose-Einstein condensate can be represented
by Fock states, Bouyer and Kasevich, as well as Dowling, 
have shown that the quantum noise
in atom interferometry using dual Bose-Einstein condensates can also
be reduced to the Heisenberg limit \cite{bouyer97,dowling98,dunn02}.

\subsection{Maximally entangled states}

The third category of states is given by the maximally
entangled states. It
was shown by Wineland and co-workers \cite{wineland96}
that the optimal
frequency measurement can be achieved by using {\it maximally
entangled states}, 
which have the following form: 
\begin{equation}\label{wineland}
 |\varphi_N\rangle = \frac{1}{\sqrt{2}}\left( \left|N,0 \right\ra_{AB}
   + \left|0,N \right\ra_{AB}\right)\; .
\end{equation}
This state has an immediate resemblence with the state in
Eq.~(\ref{cosn})
after acquiring a phase shift of $e^{iN\varphi}$.

In terms of quantum logic gates, the maximally correlated state of the
form of Eq.~(\ref{wineland}) can be made using a Hadamard gate and a
sequence of C-NOT gates.  
Note that one distinctive feature, compared to the other schemes described above,
is that the state of the form Eq.\ (\ref{wineland})
is the desired quantum state after the first beam splitter
in the Mach-Zehnder interferometer, not the input state as discussed in Sec.\ 2. 
In that the desired input state is described as the inverse beam-splitter
operation to the state of Eq.~(\ref{wineland}). 

All the interferometric schemes using entangled or dual-Fock input
states show a sensitivity approaching $1/N$ only asymptotically. 
However, using the maximally correlated states of Eq.~(\ref{wineland}),
the phase sensitivity is equal to $1/N$, even for a small $N$.
On the other hand, for atom interferometers using Bose-Einstein condensates, 
it is recently pointed out that the dual Fock state
is superior to the maximally correlated states when losses are present \cite{dunn02}.

\section{Quantum interferometric lithography}\label{litho}

Quantum correlations can also be applied to optical lithography. 
In recent work it has been shown that the Rayleigh diffraction limit
in optical lithography can be circumvented by the use of
path-entangled photon number states \cite{boto00}. The desired
$N$-photon path-entangled state, for $N$-fold resolution enhancement,
is again of the form given in Eqs.~(\ref{entang}) and (\ref{wineland}).

Consider the simple case of a two-photon Fock state $|1\ra_A |1\ra_B$,
which is a natural component of a spontaneous parametric
down-conversion event. After passing through a 50/50 beam splitter,
it becomes an entangled number state of the form $|2\ra_A |0\ra_B +
|0\ra_A |2\ra_B$. Quantum interference suppresses the probability amplitude 
of $|1\ra_A |1\ra_B$. According to quantum mechanics, it is not
possible to tell whether both photons took path $A$ or $B$ after the
beam splitter. 

When parametrizing the position $x$ on the surface by $\varphi=\pi
x/\lambda$, the deposition rate of the two photons onto the substrate
becomes $1+\cos 2\varphi$, which has twice better resolution
$\lambda/8$ than that of single-photon absorption, $1 + \cos\varphi$,
or that of uncorrelated two-photon absorption, $(1 +\cos\varphi)^2$.
For $N$-photon path-entangled state of Eq.~(\ref{wineland}), we obtain
the deposition rate $1+\cos N\varphi$, corresponding to a resolution
enhancement of $\lambda/(4N)$. 

It is well known that the two-photon path-entangled state of
Eq.~(\ref{wineland}) can be generated using a Hong-Ou-Mandel (HOM)
interferometer \cite{hom87} and two single-photon input states.
A 50/50 beam splitter, however, is not sufficient for producing
path-entangled states with a photon number larger than two
\cite{campos89}. 

The maximally correlated state of the
form of Eq.~(\ref{wineland}) can be made using a Hadamard gate and a
sequence of $N$ consecutive C-NOT gates. However, building 
optical C-NOT gates normally
requires  large optical nonlinearities. 
Consequently, in generating such states
it is commonly assumed that large $\chi^{(3)}$ nonlinear 
optical components are needed for $N >2$.
Knill, Laflamme, and Milburn proposed a method for creating
probabilistic single-photon quantum logic gates based on
teleportation. The only resources for this method are linear optics
and projective measurements \cite{klm01}. Probabilistic quantum
logic gates using polarization degrees of freedom have been demonstrated by
Franson and co-workers \cite{pittman01,pittman02}.
These works make linear optics a viable candidate for quantum computing.  
Using the concept of projective measurements, we have previously
demonstrated that the desired
path-entangled states can be created
when conditioned on the measurement outcome \cite{lee02a,kok02a}.

\section{Conclusion}

There are certain optical decoherence processes that will serve 
to prevent sensors from operating at the $1/N$  Heisenberg limit. 
Chief among these are noise terms in the nonlinear photon production 
and random photon loss from the interferometer in transit, 
due to the inefficiency of the detectors. 
We now begin to see the relationship between the problems 
of quantum computing and that of Heisenberg-limited interferometry. 
In the case of quantum computing, interference and quantum entanglement 
are used to compute a result that can not be obtained on a classical computer. 
In the case of the quantum sensor, interference and quantum entanglement 
are used to obtain a sensitivity
that can not be gotten on a classical device. 
In both cases, decoherence in the actual physical device introduces errors 
that degrade the outcome of the measurement away from the desired ideal result.
What is different between these two fields, up until now, 
is that quantum error correction schemes and 
circuits have been developed for quantum computers to 
overcome the losses to decoherence, 
but such schemes have yet to be worked out for the interferometric 
setting of the quantum sensor-interferometer. 

Quantum computing in its current state of infancy presents two extremes: 
mind-boggling computational potential, which requires daunting 
theoretical and engineering problems to be solved \cite{dowling03}. 
These two philosophical poles seem to also be present in quantum sensing, 
but to a much lesser extreme. 
For instance, the quantum-computing problem of factoring a 200-digit number 
seems to require a large-scale, universal, programmable, quantum-digital computer. 
However, the simple quantum error-correcting circuits 
needed to obtain improved sensors may turn out to be more technologically 
feasible ``analog'' quantum devices that carry out a much simpler task. 
Since the quantum sensor requires highly entangled and correlated particles 
for the input state, 
this technological problem has many of the same difficult drawbacks 
in making a universal digital quantum computer, but at a much smaller analog scale. 
Even in the presence of quantum noise and decoherence, 
the recent simulations
indicate a substantial improvement over classical sensors can be 
obtained---at least for a particular choice of correlated input states \cite{kim98,dunn02}. 
The challenge then is to see if quantum error correction and coherence protection 
can be implemented on top of these or similar schemes in order to approach 
the 1/N quantum Heisenberg limit.

\section*{Acknowledgments}
J.P.D. would like to dedicate this paper to the memory of his Ph.D. advisor, mentor, and friend -- the late Asim O. Barut -- on this ten-year anniversary of his death.
Part of this work was carried out (by K.T.K.)
at the Jet Propulsion Laboratory under 
a contract with the National Aeronautics and Space Administration (NASA). 
K.T.K. acknowledges support from the National Research Council and
NASA, Codes Y and S. J.P.D. and H.L. acknowledge support from the Horace C. Hearne Jr. Foundation, the Advanced  Research and Development Activity, The Army Research Office, the National Security Agency, and the National Reconnaissance Office.

%%%%%%%%%%%%%%%%%%%%%%%%%%%%%%%%%%%%%%%%%%%%%%%%%%%%%%%%%%%%%
%%%%% References %%%%%

\end{document}